\documentclass[aps,preprint,groupedaddress,nofootinbib,,longbibliography]{revtex4-1}
\usepackage{amssymb}
\usepackage{amsmath}
\usepackage{txfonts}
\usepackage{epsfig}
\usepackage{caption2}
\usepackage{subfigure}
\usepackage{float}
\usepackage{indentfirst}
\usepackage[breaklinks,colorlinks,citecolor=green,urlcolor=magenta,linkcolor=red]{hyperref}
\usepackage{graphicx}
\usepackage{epstopdf}
\usepackage{diagbox}
\usepackage{threeparttable}
\usepackage{dcolumn}
\usepackage{array}
\usepackage{booktabs}
\usepackage{multirow}
\usepackage{makecell}

\begin{document}

\title{\boldmath\ Forward-backward asymmetry induced $CP$ asymmetry of $B^{\pm}\rightarrow \pi^{\pm}\pi^{+}\pi^{-}$}
\author{Ya-Rui Wei}
\author{Zhen-Hua Zhang}
\address{School of Nuclear Science and Technology, \\University of South China, Hengyang, Hunan, 421001, China.}
\date{\today}

\begin{abstract}
$CP$ violation of the decay $B^{\pm}\rightarrow \pi^{\pm}\pi^{+}\pi^{-}$ in the $f_{0}(500)-\rho(770)^{0}$ interfering region is analyzed.
The forward-backward asymmetries (FBAs) and the corresponding $CP$ asymmetries FB-$CP$As are particularly investigated.
To isolate the $CP$V caused by the interference of different partial wave more cleanly, we also introduce the direct-$CP$V-subtracted FB-$CP$A.
Based on the LHCb data, we extract the FBAs, FB-$CP$As,  direct-$CP$V-subtracted FB-$CP$A, as well as the regional $CP$As with invariant mass of the $\pi^{+}\pi^{-}$ pair in the range 0.2 GeV/$c^{2}<\sqrt{s_{\text{low}}}<1.8$ GeV/$c^{2}$.
It is found that the (direct-$CP$V-subtracted) FB-$CP$As are quite large in the $f_{0}(500)-\rho(770)^{0}$ interfering region, which confirms that the interference of the intermediate resonances $f_{0}(500)$ and $\rho(770)^{0}$ plays an important role for the $CP$ violation of the three-body decay channel $B^{\pm}\rightarrow \pi^{\pm}\pi^{+}\pi^{-}$.
\end{abstract}

\maketitle
\section{\label{sc:introduction}Introduction}
Charge-parity ($CP$) violation was discovered by J.W.Cronion and V.L.Fitch in the neutral kaon system in 1964 \cite{Christenson:1964fg}.
It is closely related to the  matter-antimatter symmetry in our universe \cite{Sakharov:1967dj}, and is one of the most basic and important properties of the weak interaction.
In the Standard Model (SM), $CP$ violation ($CP$V) results from the weak complex phase in the Cabibbo-Kobayashi-Maskawa (CKM) matrix that reflects the transitions of different generations of quarks \cite{Cabibbo:1963yz,Kobayashi:1973fv}.
To date, $CP$V has been observed in the $K$, $B$ and $D$ meson systems \cite{Christenson:1964fg,KTeV:1999kad,BaBar:2001pki,Belle:2001zzw,Belle:2010xyn,BaBar:2010hvw,LHCb:2013syl,LHCb:2019hro}, all of which are consistent with the KM mechanism of SM \textcolor{blue}{\cite{Kobayashi:1973fv}}.

The study on $CP$V in multi-body decays of the bottom and the charmed hadrons plays an increasingly important role in both testing the KM mechanism of SM and looking for new sources of $CP$V.
Interestingly, large $CP$As regionalized in part of the phase space in some three-body decay channels of $B$ mesons have been reported by LHCb \cite{LHCb:2013ptu,LHCb:2014mir,LHCb:2019sus,LHCb:2022nyw}.
Meanwhile, the integrated $CP$As are relatively small due to the cancellation among different parts of the phase space.
Among these three-body decay channels, $B^{\pm}\rightarrow \pi^{\pm}\pi^{+}\pi^{-}$ is one of the most extensively studied.
Amplitude analysis of this decay channel shows that  $\rho(770)^{0}$ is the dominant resonance \cite{CLEO:2000xjz,Belle:2002ezq,BaBar:2005jqu,BaBar:2009vfr}.
The large regional $CP$A observed by LHCb located right in part of the $\rho(770)^{0}$ region where the angle (in the rest frame of $\rho(770)^{0}$) of the two pions with the same charge with $B$---which will be denoted as $\theta$ in this paper---is smaller that $90^{0}$ \cite{LHCb:2014mir,LHCb:2019sus}.
Theoretical analysis indicates that the aforementioned large regional $CP$A can be explained by the interference of $\rho(770)^{0}$ with the nearby resonance $f_{0}(500)$, where the corresponding amplitudes are respectively $P$- and $S$-waves \cite{Zhang:2013oqa,Dedonder:2010fg,AlvarengaNogueira:2015wpj,Bediaga:2015mia,Zhang:2015xea,Cheng:2020iwk}.
The interference behavior around $\rho(770)^{0}$ can be well explained based on a QCD factorization approach for the weak amplitude of $B^{-}\rightarrow \rho(770)^{0}\pi^{-}$ and $B^{-}\rightarrow f_{0}(500)\pi^{-}$ \cite{Cheng:2020ipp}.

Although it can be well explained by the interference of $f_{0}(500)$ and $\rho(770)^{0}$, the large regional $CP$A in $B^{\pm}\rightarrow \pi^{\pm}\pi^{+}\pi^{-}$ entangles all kinds of contributions other than the aforementioned $f_{0}(500)-\rho(770)^{0}$ interference, such as contributions from interference between the tree and penguin part corresponding to $\rho(770)^{0}$.
Recently, one of the authors (Z.H.Z.) has proposed to study $CP$V induced by the interference of the nearby resonance through the forward-backward asymmetry (FBA) of the final particle and the FBA induced $CP$ Asymmetry (FB-$CP$A) \cite{Zhang:2021zhr}.
One advantage of this method is that it can isolated $CP$V caused by the interfering effect of the nearby resonances (respectively corresponding to even- and odd-waves) \cite{Hu:2022eql}.

The large data sample allows LHCb to study the $CP$V caused by the interference of the $S$- and $P$-waves in more detail.
To do this, the event yields are allocated into bins according to $\cos\theta>0$ or $\cos\theta<0$.
In this way, the corresponding regional $CP$As can be systematics studied \cite{LHCb:2014mir}.
However, the analysis of the FBAs and the FB-$CP$As is absent in the LHCb's previous works.
This motivates us to performed the analysis of the FBAs and the FB-$CP$As of $B^{\pm}\rightarrow\pi^{\pm}\pi^{+}\pi^{-}$ in this paper, based on the LHCb's data in Refs. \cite{LHCb:2014mir} and \cite{LHCb:2019sus}.

This paper is organized as follows.
In Sec. \ref{sec:FBICPA}, we first illustrate the definitions of FBA and FB-$CP$A in detail.
In Sec. \ref{sec:LHCb analysis},
based on the data sample corresponding to an integrated luminosity of 3.0 fb$^{-1}$ collected by the LHCb detector \cite{LHCb:2014mir},
we extract the regional $CP$As, the FBAs, the FB-$CP$As, as well as the newly introduced $CP$ observables direct-$CP$V-subtracted FB-$CP$As.
A fit of $\cos\theta$-dependence of $CP$A with only the inclusion of the amplitudes corresponding to $f_{0}(500)$ and $\rho(770)^{0}$ is presented at the end of this section.
In Sec. \ref{sec:con}, we briefly give the conclusion.

\section{\label{sec:FBICPA}\boldmath\ THE FORWARD-BACKWARD ASYMMETRY INDUCED $CP$ ASYMMETRY}
\begin{figure}[H]
  \centering
  \includegraphics[width=0.7\textwidth]{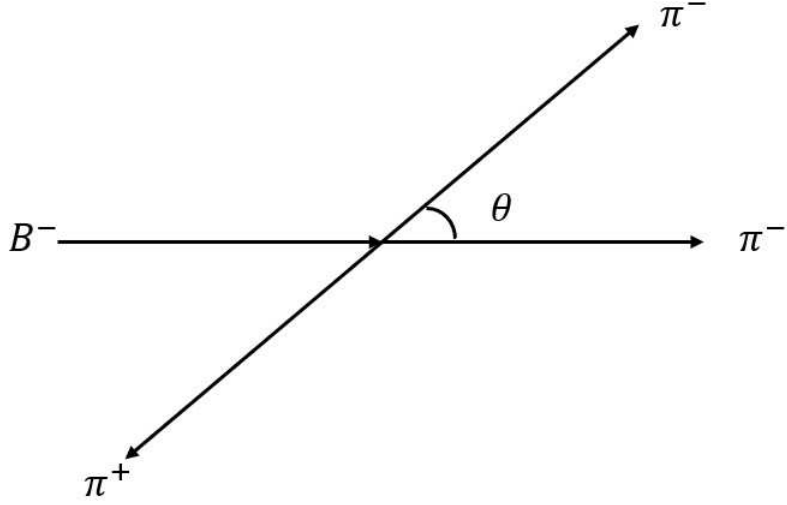}\\
  \caption{The definition of $\theta$.}
    \label{FIG:1}
\end{figure}

We first illustrate the definition of several $CP$A observables for the decay process $B^{-}\rightarrow \pi^{-}\pi^{+}\pi^{-}$.
In the c. m. frame of one of the $\pi^{+}\pi^{-}$ pair with low invariant mass,
the aforementioned angle $\theta$ between the two $\pi^{-}$'s are illustrated in Fig. \ref{FIG:1}.
In practice, the invariant mass of the $\pi^{+}\pi^{-}$ pair is separated into small intervals.
For each interval, it can be further separated into two small regions according to $\cos\theta>0$ or $\cos\theta<0$, which will be denoted as $\Omega_{i}^{+}$ or $\Omega_{i}^{-}$, respectively, where the subscript $i$ is the label of the small interval.
The regional $CP$As, which have been repeatedly dealt with in the literature both theoretically and experimentally, are defined as
\begin{align}
A_{CP}^{\Omega_{i}^{\pm}}=\frac{N_{\Omega_{i}^{\pm}}-\overline{N}_{\Omega_{i}^{\pm}}}{N_{\Omega_{i}^{\pm}}+\overline{N}_{\Omega_{i}^{\pm}}},
\end{align}
for the region $\Omega_{i}^{\pm}$, where $N_{\Omega_{i}^{\pm}}$ and $\overline{N}_{\Omega_{i}^{\pm}}$ are the event yields of $B^{-}\rightarrow\pi^{-}\pi^{-}\pi^{+}$ and $B^{+}\rightarrow\pi^{+}\pi^{+}\pi^{-}$ in the region $\Omega_{i}^{\pm}$, respectively.
Of course, the regional $CP$A for $\Omega_{i}\equiv\Omega_{i}^{+}+\Omega_{i}^{-}$ is defined as
\begin{align}
A_{CP}^{\Omega_{i}}=\frac{N_{\Omega_{i}}-\overline{N}_{\Omega_{i}}}{N_{\Omega_{i}}+\overline{N}_{\Omega_{i}}},
\end{align}
where $N_{\Omega_{i}}$ and $\overline{N}_{\Omega_{i}}$ are respectively the event yields of $B^{-}\rightarrow\pi^{-}\pi^{-}\pi^{+}$ and $B^{+}\rightarrow\pi^{+}\pi^{+}\pi^{-}$ in $\Omega_{i}$.

The FBA of the interval $i$ of $B^{-}\rightarrow\pi^{-}\pi^{+}\pi^{-}$ is defined as the relative event yields difference between $\Omega_{i}^{+}$ and $\Omega_{i}^{-}$:
\begin{align}
A_{i}^{FB}\equiv\frac{N_{\Omega_{i}^{+}}-N_{\Omega_{i}^{-}}}{N_{\Omega_{i}^{+}}+N_{\Omega_{i}^{-}}}.
\end{align}
The FB-$CP$A of the interval $i$ is defined as
\begin{equation}
\begin{aligned}
A^{FB}_{CP,i}&=\frac{1}{2}\left(A_{i}^{FB}-\overline{A_{i}^{FB}}\right),
\label{eq8}
\end{aligned}
\end{equation}
where $\overline{A^{FB}}$ represents the FBA of the interval $i$ for $B^{+}\rightarrow\pi^{+}\pi^{+}\pi^{-}$.
Comparing with the regional $CP$As, the FB-$CP$A is free from the assumption of equal production of $B^{-}$ and $B^{+}$ \cite{LHCb:2017bdt}, which reduces the corresponding systematic uncertainties.

Alternatively, one can define the $CP$A corresponding to FBA as
$A_{CP,i}^{FB,\text{alt.}}=\frac{A_{i}^{FB}-\overline{A_{i}^{FB}} }{A_{i}^{FB}+\overline{A_{i}^{FB}}}$,
which is similar with the $CP$As defined by the decay width.
However, there are good reasons for us not to use this definition here.
Mathematically, since neither $A^{FB}$ nor $\overline{A^{FB}}$ are  positive-definite, the $CP$A defined in the above equation is not bounded in $(-1, 1)$.
In other words, $A_{CP,i}^{FB}$ is normalized, while $A_{CP,i}^{FB,\text{alt.}}$ is not.
To be more specific, $A_{CP,i}^{FB,\text{alt.}}$ is in fact the relative FB-$CP$A with respect to the $CP$-averaged FBA, $A_{i}^{FB, CP-\text{av.}}\equiv\frac{1}{2}(A_{i}^{FB}+\overline{A_{i}^{FB}})$, which can be clearly seen from a transformed expression: $A_{CP,i}^{FB,\text{alt.}}={A_{CP,i}^{FB}}/{A_{i}^{FB, CP-\text{av.}}}$.
Hence the aforementioned alternative definition is questionable when one wants to make a comparison with other $CP$As, such as the regional ones.
From this perspective, the definition of Eq. (\ref{eq8}) is more reasonable for the usage in this paper
\footnote{Similar story happens in other cases, such as the hyperon decays, where the $CP$A corresponding to the decay parameter $\alpha$ are defined in the literature as $A_{CP}^{\alpha}=\frac{\alpha+\bar{\alpha}}{\alpha-\bar{\alpha}}$, while an alternative definition which is similar with Eq. (\ref{eq8}) was presented in Ref. \cite{Zhang:2022emj}.
According to the logic here, the former should be view as the relative $\alpha$-induced $CP$A with respect to the $CP$-averaged decay parameter $\alpha^{CP-\text{av.}}\equiv\frac{1}{2}(\alpha-\bar{\alpha})$, while the latter is the $\alpha$-induced $CP$A: $A_{CP}^{\alpha-\text{ind.}}\equiv\frac{1}{2}(\alpha+\bar{\alpha})$.
Of course, we are not saying that the latter definition of $CP$A is better.
On the contrary, despite non-normalized, the former does have some distinct advantages.
For example, the former defined relative $CP$A can be measured through $\Lambda_{c}^{+}\rightarrow \Lambda (\rightarrow p \pi^{-}) h^{+}$ \cite{Wang:2022tcm,Belle:2022uod} based only on the $CP$ symmetry assumption for the decay $\Lambda_{c}^{+}\rightarrow\Lambda h^{+}$ ($h^{+}=K^{+}$ or $\pi^{+}$), while the latter cannot. }.

The nonzero of FBA is caused by the interference of the odd- and even-waves \cite{Hu:2022eql}.
To see this, one express the decay amplitude as
\begin{equation}
\begin{aligned}
A&=\sum\limits_{l}a_{l}P_{l}(\cos\theta),
\label{eq11}
\end{aligned}
\end{equation}
the FBA is then
\begin{equation}
\begin{aligned}
A^{FB}&=\frac{1}{\sum\limits_{j}\left[\langle|a_{j}|^{2}\rangle/(2j+1)\right]}\mathop{\sum\limits_{\text{even}\;l}}\limits_{\mathop{\text{odd}\;k}}f_{lk}\mathfrak{R}\left(\langle a_{l}a_{k}^{*}\rangle\right),
\label{eq12}
\end{aligned}
\end{equation}
where $f_{lk}=\frac{(-)^{(l+k+1)/2}l!k!}{2^{l+k-1}(l-k)(l+k+1)[(l/2)!]^{2}\{[(k-1)/2]!\}^{2}}$ \cite{Byerly}. Consequently, the FB-$CP$A provide an effective procedure to isolate $CP$V corresponding to the interference of odd- and  even-waves. Strictly speaking, however, the FB-$CP$A contains also $CP$Vs corresponding to the difference between the decay width of $CP$-conjugate processes.
This can been seen from the denominator of the above equation, which is proportional to the decay width.
Hence, the difference between the decay width of the $CP$-conjugate processes, which is in fact the origin of the direct $CP$V, can also contribute to FB-$CP$A.
To eliminate this, one can introduce an observable, which will be called as the direct-$CP$V-subtracted FB-$CP$A, taking the form
\begin{equation}
\begin{aligned}
\tilde{A}_{CP}^{FB}&\equiv\frac{\mathop{\sum\limits_{\text{even}\;l}}\limits_{\mathop{\text{odd}\;k}}f_{lk}\mathfrak{R}\left(\langle a_{l}a_{k}^{*}\rangle-\langle \bar{a}_{l}\bar{a}_{k}^{*}\rangle\right)}{\sum\limits_{j}\left[\langle|a_{j}|^{2}\rangle/(2j+1)\right]+\sum\limits_{j}\left[\langle|\bar{a}_{j}|^{2}\rangle/(2j+1)\right]}.
\label{eq13}
\end{aligned}
\end{equation}
For the current situation, the direct-$CP$V-subtracted FB-$CP$A of the interval $i$ can be expressed as
\begin{equation}
\begin{aligned}
\tilde{A}_{CP,i}^{FB}&=\frac{\left(N_{\Omega_{i}^{+}}-N_{\Omega_{i}^{-}}\right)-\left(\overline{N}_{\Omega_{i}^{+}}-\overline{N}_{\Omega_{i}^{-}}\right)}{N_{\Omega_{i}}+\overline{N}_{\Omega_{i}}},
\label{eq14}
\end{aligned}
\end{equation}
based on the assumption that the $B^{-}$ and $B^{+}$ are equally produced.

\section{\label{sec:LHCb analysis}\boldmath\ FBA AND FB-$CP$A ANALYSIS BASED ON LHCb DATA IN $B^{\pm}\rightarrow\pi^{\pm}\pi^{+}\pi^{-}$}
\begin{figure}[H]
  \centering
  \includegraphics[width=1.0\textwidth]{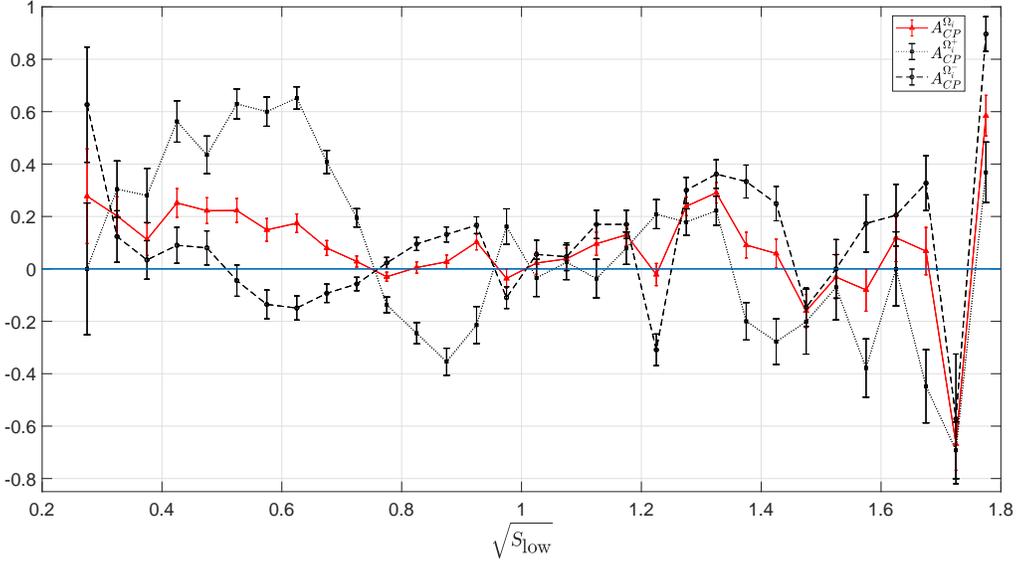}\\
  \caption{The regional $CP$As, $A_{CP}^{\Omega_{i}^{+}}$, and $A_{CP}^{\Omega_{i}^{-}}$, and $A_{CP}^{\Omega_{i}}$ of the decay channel $B^{\pm}\rightarrow\pi^{\pm}\pi^{+}\pi^{-}$ extracted from the LHCb data in Ref. \cite{LHCb:2014mir} for $\sqrt{s_{\text{low}}}$ from 0.2 GeV/$c^{2}$ to 1.8 GeV/$c^{2}$.
  The dotted, dashed, and solid lines are $A_{CP}^{\Omega_{i}^{+}}$, $A_{CP}^{\Omega_{i}^{-}}$, and $A_{CP}^{\Omega_{i}}$, respectively.}
    \label{FIG:2}
\end{figure}

\begin{figure}[H]
  \centering
  \includegraphics[width=1.0\textwidth]{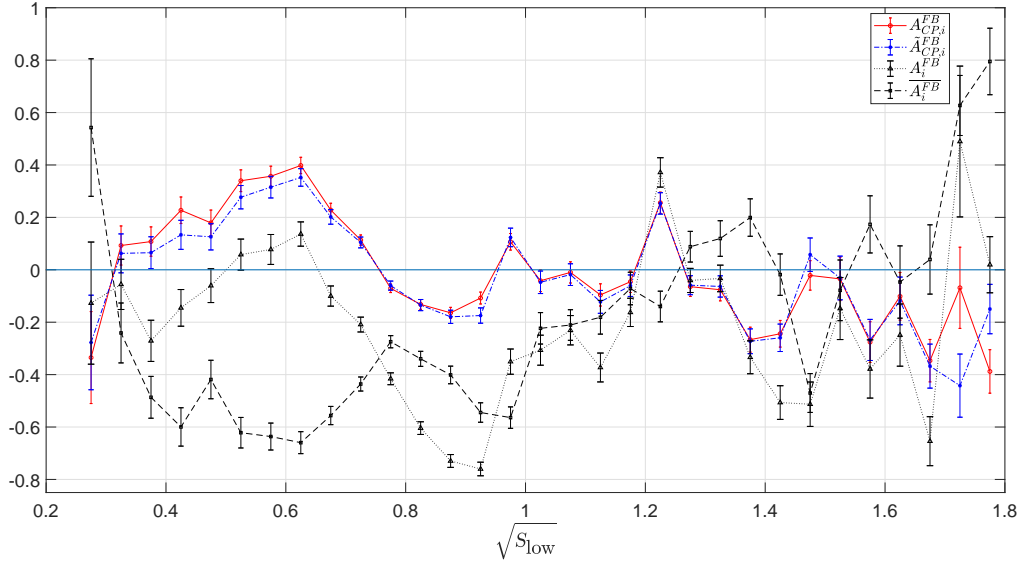}\\
  \caption{The FBAs ($A_{i}^{FB}$), the FB-$CP$As ($A_{CP,i}^{FB}$), and the direct-$CP$V-subtracted FB-$CP$As ($\tilde{A}_{CP,i}^{FB}$) of the decay channel $B^{\pm}\rightarrow\pi^{\pm}\pi^{+}\pi^{-}$ for $\sqrt{s_{\text{low}}}$ from 0.2 GeV/$c^{2}$ to 1.8 GeV/$c^{2}$.
  The dotted and dashed lines are $A_{i}^{FB}$ and $\overline{A_{i}^{FB}}$, respectively.
  The solid and dash-dotted lines are $A_{CP,i}^{FB}$ and $\tilde{A}_{CP,i}^{FB}$, respectively.}
    \label{FIG:3}
\end{figure}

Detailed analysis of the event distributions of the decay $B^{\pm}\rightarrow\pi^{\pm}\pi^{+}\pi^{-}$ has been performed by LHCb based on a data sample corresponding to an integrated luminosity of 3.0 fb$^{-1}$ \cite{LHCb:2014mir,LHCb:2019sus}.
Based on the data in Ref. \cite{LHCb:2014mir}, we can extract the regional $CP$As, the FBAs, the FB-$CP$As, as well as the direct-$CP$V-subtracted FB-$CP$As in a wide region of the $\pi^{+}\pi^{-}$ pair with lower invariant mass: 0.2 GeV/$c^{2}<\sqrt{s_{\text low}}<$1.8 GeV/$c^{2}$, which are presented in Figs. \ref{FIG:2} and \ref{FIG:3}.

FIG. \ref{FIG:2} shows the corresponding regional $CP$As of each bins with width of 0.05 GeV/$c^{2}$ lying in the range 0.2 GeV/$c^{2}<\sqrt{s_{\text{low}}}<1.8$ GeV/$c^{2}$,
while FIG. \ref{FIG:3} shows the FBAs, the FB-$CP$As, and the direct-$CP$V-subtracted FB-$CP$As.
The errors are estimated with only the inclusion of the statistical uncertainties of the event yields estimated according to $\sqrt{N}$.
Both of the two figures show interesting behavior around the $f_{0}(500)-\rho(770)^{0}$ interference region, i.e., 0.45 GeV/$c^{2}<\sqrt{s_{\text{low}}}<0.75$ GeV/$c^{2}$.

\begin{figure}[H]
  \centering
  \includegraphics[width=1.0\textwidth]{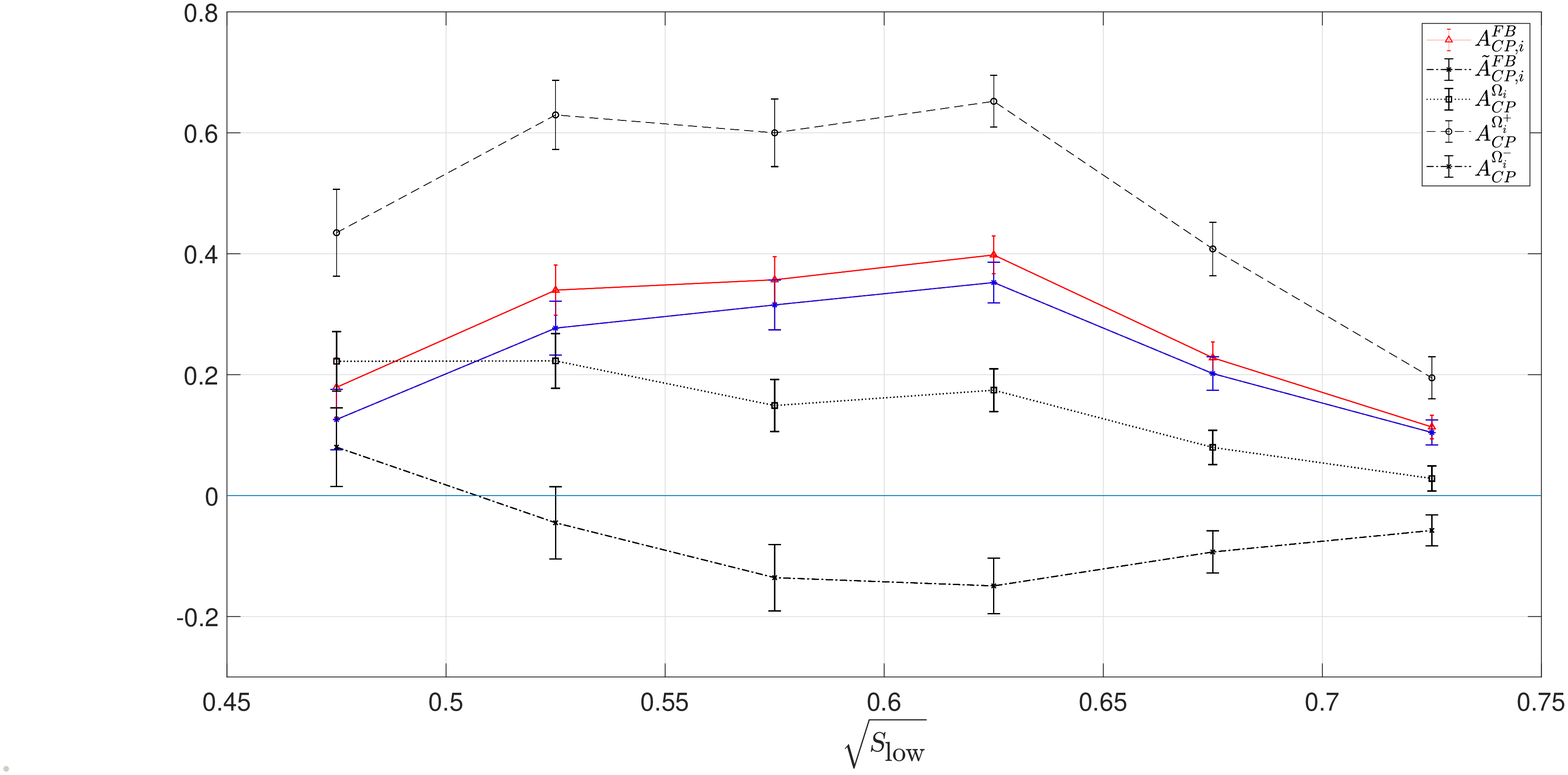}\\
  \caption{The $CP$As, $A_{CP,i}^{FB}$, $\tilde{A}_{CP,i}^{FB}$, $A_{CP}^{\Omega_{i}}$, $A_{CP}^{\Omega_{i}^{+}}$, and $A_{CP}^{\Omega_{i}^{-}}$ of the decay channel $B^{\pm}\rightarrow\pi^{\pm}\pi^{+}\pi^{-}$ for $\sqrt{s_{\text{low}}}$ from 0.45 GeV/$c^{2}$ to 0.75 GeV/$c^{2}$.
  The upper and lower solid lines are $A_{CP,i}^{FB}$ and $\tilde{A}_{CP,i}^{FB}$, respectively.
  The dotted, dashed, and dash-dotted lines are $A_{CP}^{\Omega_{i}}$, $A_{CP}^{\Omega_{i}^{+}}$, and $A_{CP}^{\Omega_{i}^{-}}$, respectively.}
    \label{FIG:4}
\end{figure}

One can see from Fig. \ref{FIG:2} that the regional $CP$As $A_{CP}^{\Omega_{i}^{+}}$ are quite large in the $f_{0}(500)-\rho(770)^{0}$ interference region.
For the FBAs, the FB-$CP$As, and the direct-$CP$V-subtracted FB-$CP$As in Fig. \ref{FIG:3}, one can see that there are big differences between $A_{i}^{FB}$ and $\overline{A_{i}^{FB}}$, resulting in a large $A_{CP,i}^{FB}$.
For a more detailed and transparent comparison, we present all the five $CP$As, $A_{CP}^{\Omega_{i}}$, $A_{CP}^{\Omega_{i}^{+}}$, $A_{CP}^{\Omega_{i}^{-}}$, $A_{CP,i}^{FB}$, and $\tilde{A}_{CP,i}^{FB}$ in Fig. \ref{FIG:4}.

The first thing one can see from Fig. \ref{FIG:4} is that $A_{CP}^{\Omega_{i}^{+}}$, $A_{CP,i}^{FB}$, and $\tilde{A}_{CP,i}^{FB}$ are quite large throughout the whole $f_{0}(500)-\rho(770)^{0}$ interference region.
Besides, the differences between $A_{CP}^{\Omega_{i}^{+}}$ and $A_{CP}^{\Omega_{i}^{-}}$ are very large too.
However, when summed up the event yields of the region $\Omega_{i}$, the resulting $CP$As $A_{CP}^{\Omega_{i}}$ are much smaller.
This means that the $CP$ violation in this region is dominated by the interference of $S$- and $P$-waves amplitudes, i.e., the interference between the amplitudes corresponding to $f_{0}(500)$ and $\rho(770)^{0}$ respectively.
This conclusion can also be verified from the small difference between $A_{CP,i}^{FB}$ and $\tilde{A}_{CP,i}^{FB}$, since their difference reflects the $CP$V {\it within} the $S$- or $P$-waves.

 \begin{table}[h!]
\centering
\newcolumntype{d}{D{.}{.}{4}}
\caption{\label{tab:4} The averaged $CP$As, $A_{CP}^{FB}$, $\tilde{A}_{CP}^{FB}$, $A_{CP}^{\Omega}$, $A_{CP}^{\Omega^{+}}$ and $A_{CP}^{\Omega^{-}}$ of the whole $f_{0}(500)-\rho(770)^{0}$ interference region, for 0.45 GeV/$c^{2}<\sqrt{s_{\text{low}}}<0.75$ GeV/$c^{2}$, where the uncertainties are statistical only.}
\begin{tabular}{|c|c|c|c|c|}
\hline
$A_{CP}^{FB}$&$\tilde{A}_{CP}^{FB}$&$A_{CP}^{\Omega}$&$A_{CP}^{\Omega^{+}}$ &$A_{CP}^{\Omega^{-}}$ \\ [0.7ex]
\hline
$0.224\pm0.012$&$0.194\pm0.013$&$0.099\pm0.013$&$0.405\pm0.020$&$-0.074\pm0.017$\\
\hline
\end{tabular}
\end{table}

Amplitude analysis of this decay channel $B^{\pm}\rightarrow\pi^{-}\pi^{+}\pi^{\pm}$ showed that $\rho(770)^{0}$ plays a dominant role and that the interference of $\rho(770)^{0}$ and $f_{0}(500)$ also contributes significantly \cite{LHCb:2019sus}.
The $CP$Vs, correspondingly, get contribution from the interference of the tree and penguin parts of the $S$- or $P$-wave amplitudes, which are in fact the origin of direct $CP$As of the decay channel $B^{\pm}\rightarrow\pi^{\pm}f_{0}(500)$ or $B^{\pm}\rightarrow\pi^{\pm}\rho(770)^{0}$, and the interference between $S$- and $P$-waves amplitudes.
All of the aforementioned contributions of $CP$V present in the regional $CP$As, $A_{CP}^{\Omega_{i}^{+}}$ and $A_{CP}^{\Omega_{i}^{-}}$.
While for the region $\Omega_i$, the contribution from the $S$- and $P$-waves interfering term cancel out totally.
Meanwhile, theoretical analysis in Sec. \ref{sec:LHCb analysis} shows that the (direct-$CP$V-subtracted) FB-$CP$As contain only the contributions from the interference between the $S$- and $P$-waves amplitudes.
The large FB-$CP$As and the direct-$CP$V-subtracted FB-$CP$As in Fig. \ref{FIG:4} indicates with no doubt that the $CP$Vs corresponding to the interference between $S$- and $P$-waves amplitudes dominant in the $f_{0}(500)-\rho(770)^{0}$ interference region.
This can also be seen from the numerical values of $A_{CP}^{FB}$, $\tilde{A}_{CP}^{FB}$, $A_{CP}^{\Omega}$, $A_{CP}^{\Omega^{+}}$ and $A_{CP}^{\Omega^{-}}$, of the whole region presented in Table. \ref{tab:4},
from which one can see that the $A_{CP}^{FB}$ and $\tilde{A}_{CP}^{FB}$ are much larger than $A_{CP}^{\Omega}$.

\begin{figure}[H]
  \centering
  \includegraphics[width=0.8\textwidth]{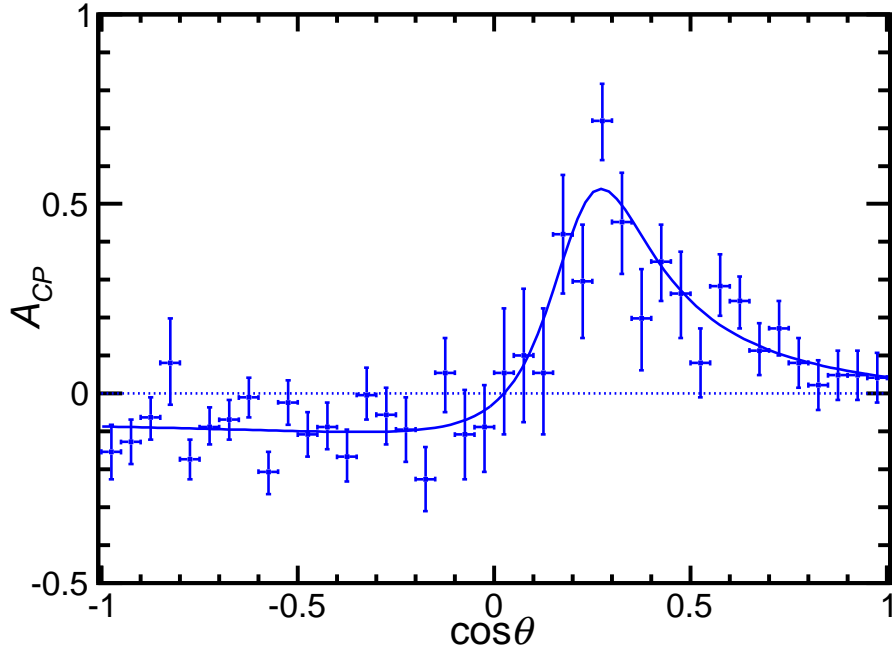}\\
  \caption{Dependence of regional $CP$As on $\cos\theta$ for 0.62 GeV/$c^{2}<\sqrt{s_{\text{low}}}<$0.78 GeV/$c^{2}$, where $\cos\theta$ is divided into 30 bins.
  The fitted curve is also shown in this figure.}
    \label{FIG:5}
\end{figure}

The interference of the $S$- and $P$-waves can also explain the fine structures of the $\cos\theta$-dependencies of the regional $CP$As.
Figure \ref{FIG:5} presents the $\cos\theta$-dependencies of the regional $CP$As
measured by LHCb for 0.62 GeV/$c^{2}<\sqrt{s_{\text{low}}}<0.78$ GeV/$c^{2}$, where $\cos\theta$ is equally divided into $30$ bins \cite{LHCb:2019sus}.
From Fig. \ref{FIG:5} one can clearly see that the regional $CP$As depend on $\cos\theta$ strongly.
It has been pointed out that the tendency of the regional $CP$As of taking opposite signs for $\cos\theta>0$ and $\cos\theta<0$ is due to the interference of $f_{0}(500)$ and $\rho(770)^{0}$ \cite{LHCb:2014mir}.
Indeed, one can fit the data in Fig. \ref{FIG:5} quite well by the inclusion of only $f_{0}(500)$ and $\rho(770)^{0}$.
The fitted curve is also shown in Fig. \ref{FIG:5}.

To fit the data, the decay amplitude is parametrized as
\begin{equation}
\begin{aligned}
\mathcal{A}_{B^{-}\rightarrow \pi^{-}\pi^{+}\pi^{-}}&=\cos\theta \mathcal{R}_{1}\frac{c_{\rho}e^{i\delta_{\rho}}}{s_{\text{low}}-m_{\rho}^{2}+im_{\rho}\Gamma_{\rho}}+\mathcal{R}_{2}\frac{c_{f}e^{i\delta_{f}}}{s_{\text{low}}-m_{f}^{2}+im_{f}\Gamma_{f}},
\label{eq9}
\end{aligned}
\end{equation}
\begin{equation}
\begin{aligned}
\mathcal{\overline{A}_{B^{+}\rightarrow \pi^{-}\pi^{+}\pi^{+}}}&=\cos\theta \mathcal{R}_{1}\frac{\bar{c}_{\rho}e^{i\bar{\delta}_{\rho}}}{s_{\text{low}}-m_{\rho}^{2}+im_{\rho}\Gamma_{\rho}}+\mathcal{R}_{2}\frac{\bar{c}_{f}e^{i\bar{\delta}_{f}}}{s_{\text{low}}-m_{f}^{2}+im_{f}\Gamma_{f}},
\label{eq10}
\end{aligned}
\end{equation}
where $c_{i}$ and $\delta_{i}$ ($i=\rho, f$) represent the corresponding amplitudes and the relative phases contribution of component $\rho(770)^{0}$ and $f_{0}(500)$, respectively, $\mathcal{R}_{1}=\sqrt{s_{\text{low}}-m_{\pi}^{2}}\sqrt{\frac{(m_{B}^{2}-s_{\text{low}}+m_{\pi}^{2})^{2}}{s_{\text{low}}}-m_{\pi}^{2}}$, $\mathcal{R}_{2}=\sqrt{s^{*}-m_{\pi}^{2}}\sqrt{\frac{(m_{B}^{2}-s^{*}+m_{\pi}^{2})^{2}}{s^{*}}-m_{\pi}^{2}}$, where $s^{*}=\frac{m_{\rho}^{2}+m_{f}^{2}}{2}$ and $m_{B}$, $m_{\pi}$, $m_{\rho}$, and  $m_{f}$ are the masses of $B$, $\pi$, $\rho(770)^{0}$, and $f_{0}(500)$, respectively.
The parameters used in Eqs. (\ref{eq9}) and (\ref{eq10}) are listed in Table \ref{tab:6}, and are borrowed from Ref. \cite{Workman:2022ynf}.
Our fit is performed in a least square method.
To fit the data, the three parameters $c_{\rho}$, $\delta_{\rho}$, and $\bar{\delta}_{\rho}$ are set fixed at $c_{\rho}=1$, $\delta_{\rho}=\bar{\delta}_{\rho}=0$.
The fitting results for the rest five of the parameters are presented in Table \ref{tab:7}.
The goodness for this fit is $\chi^2/d.o.f.=1.10$, where $\chi^2$ is the Pearson's chi-square and is defined as $\chi^{2}\equiv\sum_{j=1}\frac{(O_j-E_j)^{2}}{E_j}$ with $O_j$ and $E_j$ being the observed and expected event yields of bin $j$, and $d.o.f.$ is the degrees of freedom.

\begin{table}[h!]
\centering
\newcolumntype{d}{D{.}{.}{9}}
\caption{\label{tab:6} Input parameters used in Eqs. (\ref{eq9}) and (\ref{eq10}).}
\begin{tabular}{l||c}
\hline
\hline
Parameters&Value (in GeV)\\ [0.7ex]
\hline
 $m_{B^{\pm}}$&$5.279$\\
 $m_{\pi}$&$0.139$\\
$m_{\rho(770)^{0}}$&$0.775$\\
$\Gamma_{\rho(770)^{0}}$&$0.149$\\
$m_{f_{0}(500)}$&$0.563$\\
$\Gamma_{f_{0}(500)}$&$0.350$\\
\hline
\hline
\end{tabular}
\end{table}

 \begin{table}[h!]

\centering
\newcolumntype{d}{D{.}{.}{9}}
\caption{\label{tab:7} Parameters values obtained by the fitting.}
\begin{tabular}{c|c||c|c}
\hline
\hline
Parameter&Value&Parameter&Value \\ [0.7ex]
\hline
$c_{\rho}$ & 1 & $\bar{c}_{\rho}$ & $1.05\pm0.02$ \\
$\delta_{\rho}$ & $0$ & $\bar{\delta}_{\rho}$ &$0$ \\
$c_{f}$ & $0.49\pm0.07$ & $\bar{c}_{f}$ & $0.50\pm0.06$ \\
$\delta_{f}$ & $-0.64\pm0.06$ & $\bar{\delta}_{f}$ & $-1.36\pm0.31$\\
\hline
\hline
\end{tabular}
\end{table}

\section{\label{sec:con}\boldmath SUMMARY AND CONCLUSION}
In this paper, the FBA, the FB-$CP$A, the direct-$CP$V-subtracted FB-$CP$As and the regional $CP$As for the three-body decay of $B$ meson $B^{\pm}\rightarrow \pi^{\pm}\pi^{+}\pi^{-}$ are analyzed based on the data of LHCb \cite{LHCb:2014mir,LHCb:2019sus}.
We focus on the $CP$As in the $f_{0}(500)-\rho(770)^{0}$ interfering region.
It is found that the FB-$CP$As are quite large in this region.
According to the theoretical analysis, the large FB-$CP$A and the direct-$CP$V-subtracted FB-$CP$As can be explained by the interference of the amplitudes corresponding to $f_{0}(500)$ and $\rho(770)^{0}$.
Moreover, the analysis indicates that the interference between the amplitudes of $f_{0}(500)$ and $\rho(770)^{0}$ is the main contribution to $CP$Vs in this region.
The aforementioned interference can even explain more detailed structures of $CP$Vs in this region, according to the fitting result of Fig. \ref{FIG:5}.

In conclusion, the interference of intermediate resonances can play an important role in $CP$V of three-body decay of bottom meson.
The FB-$CP$A and the direct-$CP$V-subtracted FB-$CP$A, according the theoretical analysis and the data-based analysis of this paper, are ideal observables to study $CP$Vs coursed by the interference of intermediate resonances.
\begin{acknowledgments}
This work was supported by National Natural Science Foundation of China under Grants No. 12192261, and Natural Science Foundation of Hunan Province under Grants No. 2022JJ30483.
\end{acknowledgments}
\bibliography{www}
\end{document}